\documentclass{article}

\usepackage{arxiv}

\usepackage[utf8]{inputenc} % allow utf-8 input
\usepackage[T1]{fontenc}    % use 8-bit T1 fonts
\usepackage{hyperref}       % hyperlinks
\usepackage{url}            % simple URL typesetting
\usepackage{booktabs}       % professional-quality tables
\usepackage{amsfonts}       % blackboard math symbols
\usepackage{nicefrac}       % compact symbols for 1/2, etc.
\usepackage{microtype}      % microtypography
\usepackage{lipsum}
\usepackage{amssymb}
\usepackage{amsmath}
\usepackage{graphicx}
\usepackage{caption}
\usepackage{subcaption}
\usepackage{tabularx}

\title{Cross-beam energy transfer saturation by ion trapping-induced detuning}

\author{
  K. L. Nguyen \\
  Laboratory for Laser Energetics, University of Rochester, Rochester, New York, 14623, USA\\
  Department of Physics \& Astronomy, University of Rochester, Rochester, New York, 14623, USA\\
  Los Alamos National Laboratory, Los Alamos, New Mexico, 87545, USA \\
  \texttt{kngu@lle.rochester.edu} \\
   \And
  L. Yin \\
  Los Alamos National Laboratory, Los Alamos, New Mexico, 87545, USA \\
  \texttt{lyin@lanl.gov} \\
  \And
  B. J. Albright \\
  Los Alamos National Laboratory, Los Alamos, New Mexico, 87545, USA \\
  \And
  A. M. Hansen \\
  Laboratory for Laser Energetics, University of Rochester, Rochester, New York, 14623, USA\\
  Department of Physics \& Astronomy, University of Rochester, Rochester, New York, 14623, USA\\
  \And
  D. H. Froula \\
  Laboratory for Laser Energetics, University of Rochester, Rochester, New York, 14623, USA\\
  \And
  D. Turbull \\
  Laboratory for Laser Energetics, University of Rochester, Rochester, New York, 14623, USA\\
  \And
  R. K. Follett \\
  Laboratory for Laser Energetics, University of Rochester, Rochester, New York, 14623, USA\\
  \And
  J. P. Palastro \\
  Laboratory for Laser Energetics, University of Rochester, Rochester, New York, 14623, USA\\
  \texttt{jpal@lle.rochester.edu} \\
}

\begin{document}
\maketitle

\begin{abstract}
The performance of direct-drive inertial confinement fusion implosions relies critically on the coupling of laser energy to the target plasma. Cross beam energy transfer (CBET), the resonant exchange of energy between intersecting laser beams mediated by ponderomotively driven ion-acoustic waves (IAW), inhibits this coupling by scattering light into unwanted directions. The variety of beam intersection angles and varying plasma conditions in an implosion results in IAWs with a range of phase velocities. Here we show that CBET saturates through a resonance detuning that depends on the IAW phase velocity and that results from trapping-induced modifications to the ion distribution functions. For smaller phase velocities, the modifications to the distribution functions can rapidly thermalize in the presence of mid-Z ions, leading to a blueshift in the resonant frequency. For larger phase velocities, the modifications can persist, leading to a redshift in the resonant frequency. Ultimately, these results may reveal pathways towards CBET mitigation and inform reduced models for radiation hydrodynamics codes to improve their predictive capability. 
\end{abstract}

% keywords can be removed

\section{Introduction}
Laser-driven inertial confinement fusion (ICF) experiments employ an ensemble of laser pulses to compress a capsule of deuterium-tritium fuel encased in a thin outer ablator \cite{AtzeniBook,BettiHuricane2016}. In the direct-drive approach, these pulses impinge directly on the capsule ablator, creating shocks that drive inward fuel compression and outward mass ejection \cite{CraxtonReview2015}. In the indirect-drive approach, a hohlraum (i.e., a high atomic number cannister that houses the capsule) converts the laser photons to x-rays, which provide the shocks required for compression \cite{LindlReview}. In both approaches, the presence of a low-density plasma and the overlap of many pulses creates conditions apt for the growth of laser-plasma instabilities \cite{KruerBook}.

Among these instabilities, cross-beam energy transfer (CBET), the resonant exchange of energy between intersecting laser pulses mediated by ponderomotively driven ion-acoustic waves (IAWs), has emerged as particularly troublesome \cite{RandallCBET}. CBET inhibits the performance of both direct and indirect-drive implosions by scattering light into unwanted directions \cite{SekaDDCBET,IgorDDCBET,KritcherCBET}. In direct-drive this reduces the coupling of laser energy to the capsule, while in indirect-drive it can spoil the symmetry of the x-ray illumination. Both approaches have achieved some success in mitigating CBET by using independent wavelength shifts on the beams to detune the interaction \cite{MichelLambShift,GlenzerLambShift,MoodyCBETdetune,MarozasLambShift}. More extensive mitigation, however, requires pulses with much larger bandwidth \cite{BatesCBETBW}\textemdash a technology in active development at the Laboratory for Laser Energetics (LLE) and the Naval Research Laboratory \cite{DorrerFLUX,WeaverSRRS,LehmbergSRRS}. 

Comprehensive, predictive models of CBET can guide both ongoing and future mitigation strategies and help define the expanded ICF design space that these strategies afford. Current integrated models of ICF implosions, using radiation hydrodynamic simulations, typically implement simple linear models of CBET due to the computational expense associated with more complete models. While more sophisticated models have been developed \cite{FollettRayCaustic,ArnaudRay1,ArnaudRay2,DebayleRay}, common approximations include ray-optics (i.e., speckle and diffractive effects are ignored) and a steady-state plasma response, while neglecting nonlinear processes \cite{DebyleModelAccuracy}. This is despite a mounting body of work pointing to the importance of processes such as ion trapping, stochastic heating, two-ion decay, nonlinear sound waves, and IAW breakup \cite{WilliamsIonTrap,PierreCBETSatPRL,PierreCBETSatPOP,ChapmanTwoIon,HullerSoundWaves,yin2019saturation}. Perhaps the most convincing indication comes from a number of experiments that have observed nonlinear saturation \cite{KirkwoodSat,TurnbullSat,hansen2021cross}. The most recent of these, performed on the OMEGA laser at LLE, demonstrated that, at high intensities, a drop in the power transferred from a pump to seed pulse was accompanied by ${\sim}7{\times}$ increase in the ion temperature \cite{hansen2021cross}. 

Motivated by this observation, this manuscript provides a detailed description of the underlying physics responsible for CBET saturation for conditions relevant to these experiments. Specifically, we show that depending on the phase velocity of the IAW, CBET can saturate through two types of resonance detuning, both of which result from trapping-induced modifications to the ion distribution function. For "small" IAW phase velocities, the modifications to the distribution function rapidly thermalize in the presence of mid-Z ions, blueshifting the resonant IAW frequency. For "large" IAW phase velocities, the modifications to the distribution function persist for a longer time and redshift the resonant frequency. These results, obtained using the collisional particle-in-cell code VPIC \cite{bowersVPIC}, avoid many of the pitfalls associated with the reduced models used in radiation hydrodymanics codes and provide insight into the evolution and feedback of CBET with the coronal plasma. 

The remainder of this manuscript is organized as follows: Section II presents a conceptual overview of CBET and sources of detuning that can lead to saturation. Section III describes the setup for the VPIC simulations. Section IV covers the short time scale dynamics. Sections V describes the long time scale saturation of CBET for IAWs driven with "small" and "large" phase velocities. Section VI concludes the manuscript with a summary of the results and an outlook.

\begin{figure}
\centering
\includegraphics[scale=1.]{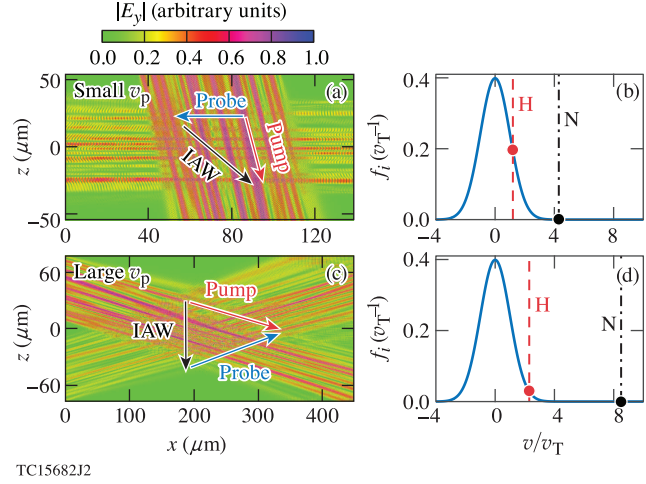}
\caption{\label{fig:wide} Electromagnetic field amplitude of the pump and probe beams for intersection angles of (a) $99^{\mathrm{o}}$ and (c) $21.4^{\mathrm{o}}$. The location of the IAW phase velocity, as determined by the plasma conditions and intersection angle, in relation to the initial distributions and thermal velocities of each ion species for the (b) $99^{\mathrm{o}}$ and (d) $21.4^{\mathrm{o}}$ geometries. In the "small" v$_p$ case (b), $\mathrm{v}_p = 1.2 \mathrm{v}_{TH} = 4.3 \mathrm{v}_{TN}$, while in the "large" case (d), $\mathrm{v}_p = 2.3 \mathrm{v}_{TH} = 8.5 \mathrm{v}_{TN}$, where $\mathrm{v}_{TH}$ and $\mathrm{v}_{TN}$ are the initial Hydrogen and Nitrogen thermal velocities.}
\label{fig:f1}
\end{figure}

\section{CONCEPTUAL OVERVIEW}
Figure \ref{fig:f1} displays the simulated CBET interaction geometries and locations of the corresponding IAW phase velocities (v$_p$) in relation to the ion distribution functions. A pump beam intersects a frequency downshifted probe beam at an angle $\theta$ in an initially uniform density plasma devoid of flow. The ponderomotive force of the two beams excites an IAW with a frequency ($\Omega$) and wavenumber ($\mathbf{k}$) determined by energy and momentum conservation: $\Omega$ = $\omega_0$ - $\omega_1$ and $\mathbf{k}$ = $\mathbf{k}_0$ - $\mathbf{k}_1$, where the subscripts 0 and 1 denote the pump and probe beams, respectively. The frequency and wavenumber, in turn, set the IAW phase velocity: $\mathrm{v}_{p}$ = $\Omega /\mathrm{k}$. 

For a given intersection angle ($\theta$), the frequency shift of the probe beam ($\Omega$), the beam intensities ($I_0$, $I_1$), and the plasma conditions determine the rate of intensity transfer from the pump to the probe beam. For plane waves with parallel polarization interacting in a linear, steady state plasma
\begin{eqnarray}
(\mathbf{v}_a\cdot\nabla)I_a = g_{ab}I_bI_a
\end{eqnarray}
where a $\neq$ b can take the values 0 or 1 to represent the pump or probe beam, $\mathbf{v}_a$ is the respective group velocity, 
\begin{eqnarray}
g_{ab} = \frac{e^2k^2\mathrm{sgn}(\omega_b-\omega_a)}{2\epsilon_{0}m_{e}^2c^2\mathrm{v}_b\omega_{b}^2k_a}\mathrm{Im}
\left[\Gamma(\Omega,\mathbf{k})\right],
\end{eqnarray}
the gain coefficient, and
\begin{eqnarray}
\Gamma(\Omega,\mathbf{k})=\frac{\chi_e(\Omega,\mathbf{k})[1+\chi_i(\Omega,\mathbf{k})]}{1+\chi_i(\Omega,\mathbf{k})+\chi_e(\Omega,\mathbf{k})},
\end{eqnarray}
the kinetic coupling factor. A summary of the derivation for Eqs. (1-3) along with the definitions for the electron and ion susceptibilities, $\chi_e$ and $\chi_i$ respectively, can be found in Appendix A.

The maximal rate of intensity transfer occurs when the IAW is resonantly excited, i.e., when $\Omega$ is close to a natural mode frequency of the plasma. Here, the linear dielectric constant, $\epsilon = 1 + \chi_e(\Omega,\mathbf{k}) + \chi_i(\Omega,\mathbf{k}) $ reaches its smallest value and $\mathrm{Im}[\Gamma(\Omega,\mathbf{k})]$ its largest value. Taylor expanding the susceptibilities about the natural mode frequency (keeping in mind that the root is complex) one finds
\begin{eqnarray}
\mathrm{Im}[\Gamma(\Omega,\mathbf{k})]\approx-\chi^2_e\Bigl(\frac{\partial \epsilon_R}{\partial \Omega}\Bigr)^{-1}
\frac{\gamma}{(\Omega-\Omega_R)^2 + \gamma^2},
\end{eqnarray}
where $\Omega_R$ is the real component of the natural mode frequency, $\gamma$ the imaginary component due to Landau damping (or growth), and $\epsilon_R = \mathrm{Re}(\epsilon)$. 

Equation (4) highlights a critical phenomenon that can contribute to the saturation of CBET: detuning. In addition to an overall $\Omega$-dependent coefficient, Eq. (4) exhibits a Lorentzian profile that peaks on resonance, i.e., when $\Omega = \Omega_R$. This resonant frequency depends on both the bulk conditions (e.g., density, flow velocity, and temperature) and detailed features (i.e., the shapes of the distribution functions) of the plasma. Suppose the probe beam has a frequency shift equal to the initial resonant frequency ($\Omega = \Omega_{R0}$). If the bulk plasma conditions or distribution functions change, the resonant frequency will evolve and detune the CBET interaction [$\Omega_R(t) \neq \Omega_{R0}$], reducing the rate of intensity transfer. In addition, $\gamma$ and the frequency dependent coefficient will also evolve, further modifying the transfer rate. All of these can occur as a result of ion heating.

In order to drive the IAW that mediates their intensity exchange, the beams must supply the IAW with energy. The IAW, in turn, can transfer this energy to the plasma ions through collisions or wave-particle interactions (e.g., wave trapping and Landau damping). Specifically, one can use the Manley-Rowe relations \cite{ManleyRowe} to find that, in steady state, the energy density of the plasma ($U$) increases as
\begin{eqnarray}
\frac{dU}{dt} = - \frac{\Omega}{\omega_0}(\hat{\mathbf{v}}_0\cdot\nabla)I_0,
\end{eqnarray}
where $\hat{\mathbf{v}}_0$ is the group velocity unit vector. For positive $\Omega$, the pump loses energy along its path, such that the right-hand-side of Eq. (5) is positive definite. Equation (5) illustrates that the CBET interaction can only occur in a quasi-steady state; plasma heating will, eventually, modify the distribution function or plasma conditions \cite{PierreCBETSatPOP,hansen2021cross}. Such a quasi-steady state picture requires that the energy density of the plasma increase slowly with respect to the IAW damping rate (i.e., $\gamma \gg |\partial_t \ln{U}| $).

In a weakly collisional plasma, such as the low-density regions where CBET is active, the initial transfer of energy from the IAW to the ions can be dominated by trapping and acceleration in the wave. These processes can cause modifications to the velocity distributions, such as flattening in the vicinity of the phase velocity and tails that extend to high velocities. Only over longer time scales will ion-ion collisions, e.g., through pitch angle scattering and slowing down, return the modified distribution to a near-equilibrium shape, albeit one with an elevated temperature and residual flow. 

The impact of and relative time scales for wave-particle interactions and collisional relaxation depend sensitively on the phase velocity of the IAW. The phase velocity not only determines the Landau damping (or growth) rate and number of ions that contribute directly to wave-particle interactions, but also how long it takes modifications at that velocity to thermalize. The source of CBET detuning, whether from the initial trapped ion modifications or their subsequent thermalization, relies on these relative time scales, and thus depends on the phase velocity of the IAW. 

Some simple examples can provide a rough idea of how these processes contribute to CBET detuning. While the natural mode frequencies for IAWs in a multi-component plasma can be somewhat complicated \cite{WilliamsIonSpecies}, for the moment consider a single or "average" ion species. Before ion-ion collisions have a chance to thermalize the modified distribution function, ion-trapping induces a nonlinear shift in the resonant frequency \cite{MoralesOneil,BergerNLfrequency}:
\begin{eqnarray}
\frac{\delta\Omega}{\Omega_R} \approx -\frac{\alpha_i}{\sqrt{2\pi}}\Bigl(\frac{\delta n_e}{n_{e0}}\Bigr)^{1/2}\Bigl(\frac{Z T_e}{T_i}\Bigr)^{3/2}(u^2-1)e^{-u^2/2},
\end{eqnarray}
where $\alpha_i$ is a numerical coefficient on the order of 1, $\delta n_{e}$ the amplitude of the IAW electron density perturbation, $n_{e0}$ the ambient electron density, $Z$ the ionization level, $T_e$ and $T_i$ the electron and ion temperatures respectively, $u = \mathrm{v}_p/\mathrm{v}_T$, $\mathrm{v}_T = \sqrt{k_bT_i/M_i}$ the thermal velocity, and $M_i$ the ion mass. The most important feature of Eq. (6) is that for typical conditions (i.e., $u>1$) trapping results in a redshift of the resonant frequency. Here $\delta n_{e} \ll n_{e0}$ has been assumed, and the electron contribution to the nonlinear frequency shift has been dropped due to its smallness for the parameters of interest (i.e., a multi-component plasma with a light ion species $u \gtrsim 1$).

As ion-ion collisions thermalize the trapping-induced modifications to the distribution function, both the flow velocity ($\mathbf{v}_F$) and the ion temperature can increase. In contrast to the trapped ion frequency shift, this blueshifts the resonant frequency:
\begin{eqnarray}
\Omega_R = \Bigl[\frac{1}{1+k^2\lambda^2_D} + \frac{3T_i}{ZT_e}\Bigr]^{1/2}kc_s + \mathbf{k}\cdot\mathbf{v}_F
\end{eqnarray}
where $\lambda_{D} = \mathrm{v}_T/\omega_{pe}$ is the Debye length, $\omega_{pe} = \sqrt{e^2n_{e0}/\epsilon_0m_e}$ the electron plasma frequency, $c_s = \sqrt{ZT_e/M_i}$ the sound speed, and the fluid approximation has been used to find the root of $\epsilon(\Omega,\mathbf{k})$. Note that, due to momentum conservation, the flow develops in the direction of $\mathbf{k}$, such that $\mathbf{k} \cdot \mathbf{v}_F > 0$. Over longer time scales, the distribution function will isotropize, and the energy in the flow will be converted to thermal energy. 

During the course of the above discussion, a number of simplifying approximations have been made in order to provide a conceptual overview of CBET and the sources of detuning that can lead to its saturation. These include plane wave pump and probe beams, steady-state plasma responses and adiabatic evolution, single or average ion species, and neglect of additional nonlinear processes that can contribute to CBET saturation or modify the underlying plasma conditions. To determine the extent of these effects and examine CBET saturation in more detail, a more comprehensive description, such as that afforded by particle-in-cell simulations, is needed.

\begin{table}
    \caption{Simulation parameters relevant to the OMEGA laser-plasma interactions experimental platform \cite{hansen2021cross}. $\Omega_{R0}$ and $\gamma$ were found by solving $\epsilon(\Omega,\mathbf{k})=0$ with the initial plasma conditions. $G_0 = \mathrm{ln}(I_1^{\mathrm{out}}/I_1^{\mathrm{in}})$ is the initial gain predicted by linear, steady state theory, including pump depletion and using an effective interaction length $L = d/\sin\theta$ (see Appendix A for details).}
        \centering
        \begin{tabular}{lcc}
        \hline \hline
        $ $&Small v$_{p}$&Large v$_{p}$\\
        \hline
        $T_e$ (eV) & $600$ & $840$\\
        $T_i$ (eV) & $150$ & $130$\\
        $n_{e0}$ $(n_{cr})$ & $0.6\%$ & $1.2\%$\\
        %$v_{ph}^{IAW}$ & $1.2v_{th,H}(4.3v_{th,N})$ & $2.3v_{th,H}(8.5v_{th,N})$\\
        $\Omega_{R0}$ (Rad/s) &$3.6{\times}10^{12}$&$1.6{\times}10^{12}$\\
        $\gamma$ $(\Omega_{R0}) $& $0.14$ & $0.18$\\
        $\theta$ $(\textrm{degrees}) $ & $99$ & $21.4$\\
        $\Delta\lambda$ (\AA) & $2.5$ & $1.1$ \\
        $G_{0}$  & $0.52$ & $1.50$\\
        \hline \hline
        \end{tabular}
    \label{tab:t1}
\end{table}

\section{SIMULATION SETUP}

Two-dimensional collisional VPIC simulations (Appendix B) \cite{bowersVPIC} were performed to model CBET saturation for parameters relevant to experiments conducted on the OMEGA laser-plasma interactions platform (Table \ref{tab:t1}) \cite{hansen2021cross}. In each of the simulated configurations [Figs. \ref{fig:f1}(a) and (c)], a speckled pump beam with vacuum wavelength $\lambda_{0}$ = 351 nm and average incident intensity $I_0^{\mathrm{in}} = 2.2\times10^{15}$ W/cm$^{2}$ intersected a speckled probe beam with an average incident intensity $I_1^{\mathrm{in}} = 5\times10^{14}$ W/cm$^{2}$. The vacuum wavelength shift of the probe beam ($\Delta\lambda = -2\pi c\Omega/\omega_0^2$) was chosen to maximize the linear gain rate ($g_{10}$) and drive the IAW near resonance ($\Omega_{R0}$). Both beams were s-polarized and had a flat-top transverse profile with a width ($d$) of 68 $\mu$m and f/6.7 speckles.

The plasma consisted of two fully ionized ion species, Hydrogen ($Z=1$) and Nitrogen ($Z=7$), and electrons. The fractional composition by number of Hydrogen (55\%) and Nitrogen (45\%) was chosen to emulate the IAW Landau damping in direct drive ICF implosions. The electrons were distributed in velocity as a super-Gaussian of order 3 as measured in the experiments \cite{hansen2021cross}. The non-Maxwellian shape results from the Langdon effect, in which collisional absorption of electromagnetic radiation predominantly heats low-velocity electrons \cite{langdon1980nonlinear,matte1988non,TurnbullLangdon,AviLangdon}. In the experiments, the Langdon effect was primarily driven by the pump beams.

\begin{figure}
\centering
\includegraphics[scale=1]{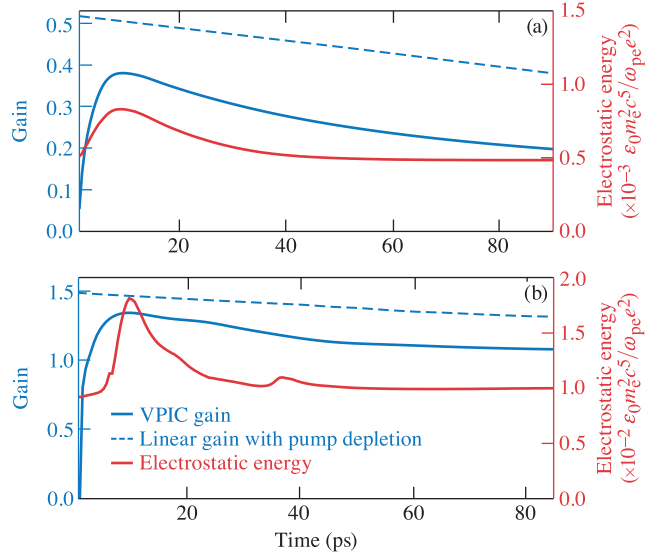}% Here is how to import EPS art
\caption{\label{fig:wide} Evolution of the CBET gain and electrostatic energy for the (a) "small" and (b) "large"  phase velocity cases. The CBET gain generally tracks the electrostatic energy. Due to an interplay of IAW transverse breakup and reduced Landau damping from ion trapping, the gain saturates at a value lower than the linear, steady state gain. The linear gain is calculated as described in Appendix A using the time-dependent ion temperatures from VPIC.}
\label{fig:f2}
\end{figure}

The two interactions displayed in Figs. \ref{fig:f1}(a) and (c), with parameters found in Table \ref{tab:t1}, drive IAWs with disparate phase velocities [Figs. \ref{fig:f1}(b) and (d)]. In the first interaction, referred to as the "small" case, $\mathrm{v}_p = 1.2 \mathrm{v}_{TH} = 4.3 \mathrm{v}_{TN}$, while in the second, or "large" case, $\mathrm{v}_p = 2.3 \mathrm{v}_{TH} = 8.5 \mathrm{v}_{TN}$, where $\mathrm{v}_{TH}$ and $\mathrm{v}_{TN}$ are the initial Hydrogen and Nitrogen thermal velocities, respectively. Naively, one may expect that the IAW phase velocity is independent of the interaction geometry and density, and depends only on $\sqrt{ZT_e}$ through the sound speed ($c_s$). However, separate calculations that solved for the roots of $\epsilon({\Omega},\mathbf{k}) = 0$ in the "small and "large" cases indicate that the primary difference in v$_p$ arises from the angle and density dependence of the $k\lambda_D$ corrections to the IAW frequency.

Aside from the IAW phase velocities, the geometry and density result in different gains for the two interactions. The initial linear, steady state gain, $G_0 = \mathrm{ln}(I_1^{\mathrm{out}}/I_1^{\mathrm{in}})$, is nearly $3\times$ higher in the large v$_p$ case due to the longer interaction length ($L=d/\sin\theta$) and higher density (note $G_0$ includes the effect of pump depletion). Nonetheless, the steady state amplitude of the IAW is nearly the same: $|\delta \hat{n}_e/n_{e0}| \sim 0.08$. As discussed in the following section, the large amplitude of the IAW affects the early-time saturation.

\begin{figure}
\centering
\includegraphics[scale=1]{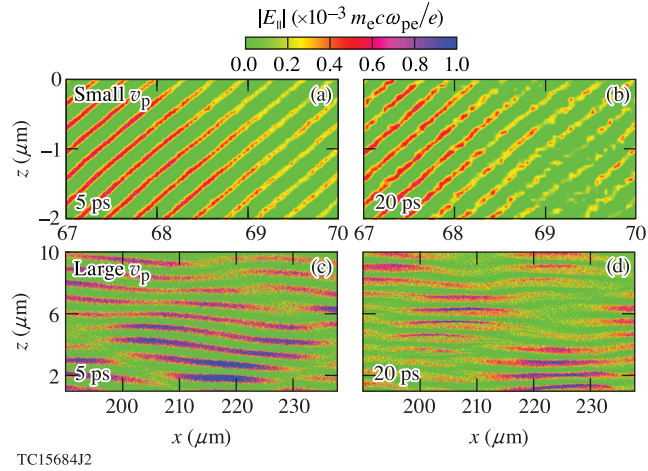}% Here is how to import EPS art
\caption{\label{fig:wide} Phase fronts of the IAW during the early time saturation stage at 5 ps and 20 ps (left and right columns, respectively) for the small and large v$_p$ cases (top and bottom rows, respectively). By 20 ps, the IAWs have undergone transverse breakup.}
\label{fig:f3}
\end{figure}

\section{EARLY TIME DYNAMICS}
Figure \ref{fig:f2} demonstrates that, in both the small and large v$_p$ cases, CBET evolves through three stages: an initial linear stage ($\lesssim 5$ ps), an early saturation stage ($\sim 5 - 20$ ps), and a final, late time saturation stage ($\gtrsim 20$ ps). During each of these stages, the gain, i.e., $G=\mathrm{ln}(P_1^{\mathrm{out}}/P_1^{\mathrm{in}})$ where $P_1^{\mathrm{in}}$ and $P_1^{\mathrm{out}}$ are the probe input and output powers, tracks the electrostatic energy. The initial stage corresponds to transient growth of the IAW \cite{DivolTransient} as the interaction attempts to evolve towards a linear, steady state. 

Before this state can be reached, however, the interaction becomes nonlinear and the IAW undergoes transverse breakup \cite{YinTPMIPRL,yin2019saturation}. As illustrated in Fig. \ref{fig:f3}, the IAW initially exhibits coherent phase fronts along $\bold{k}$, but after 20 ps the fronts have broken up into smaller transverse structures. Due to its observed correlation with ion trapping, the breakup likely results from the trapped particle modulational instability (TPMI) \cite{KruerTPI,DewarTPMI,BrunnerTPMI,YinTPMIPRL}. In the TPMI, the nonlinear frequency shift from ion trapping [as in Eq. (6)] combined with inhomogeneity in the IAW amplitude create variations in the phase velocity across the phase fronts. If a section of the phase front advances or retards by more than $\sim\pi/2$ with respect to adjacent sections, the front breaks. At this point, the wave amplitude crashes and the energy is transferred to the ions. The local dissipation of the wave prevents additional trapping and changes to the phase velocity. In fact, the rapid drop in electrostatic energy after $\sim10$ ps results from initially trapped ions carrying away the energy of the now broken IAW.  

The gain in both the small and large phase velocity cases peaks at the onset of transverse breakup. For small v$_p$, the gain only reaches $80\%$ of the linear gain calculated using the plasma conditions at that time. Here, due to the large number of H ions at v$_p$ [Fig. \ref{fig:f1}(b)], the IAW can acquire a sizable nonlinear frequency shift at an amplitude smaller than the linear, steady state prediction (i.e., $|\delta \hat{n}_e/n_{e0}|$). The nonlinear frequency shift, in turn, results in transverse breakup, which limits the average IAW amplitude and prevents CBET from reaching the predicted linear gain. This is despite the reduced Landau damping due to transient trapping and flattening of the distribution function in the surviving regions of coherent phase.

In contrast, for large v$_p$, the gain reaches $\sim 90\%$ of the linear prediction. Here, both the H and N distributions have fewer ions at v$_p$ [Fig. \ref{fig:f1}(d)]. As a result, the IAW can grow to a larger amplitude before acquiring a sizable nonlinear frequency shift and undergoing transverse breakup. While the gain approximately matches linear theory in this case, the agreement results from a near-coincidental balancing of nonlinear effects: weakened Landau damping due to trapping and a drop in the average IAW amplitude due to transverse breakup. 

\section{LATE TIME SATURATION}

\begin{figure}
\centering
\includegraphics[scale=1]{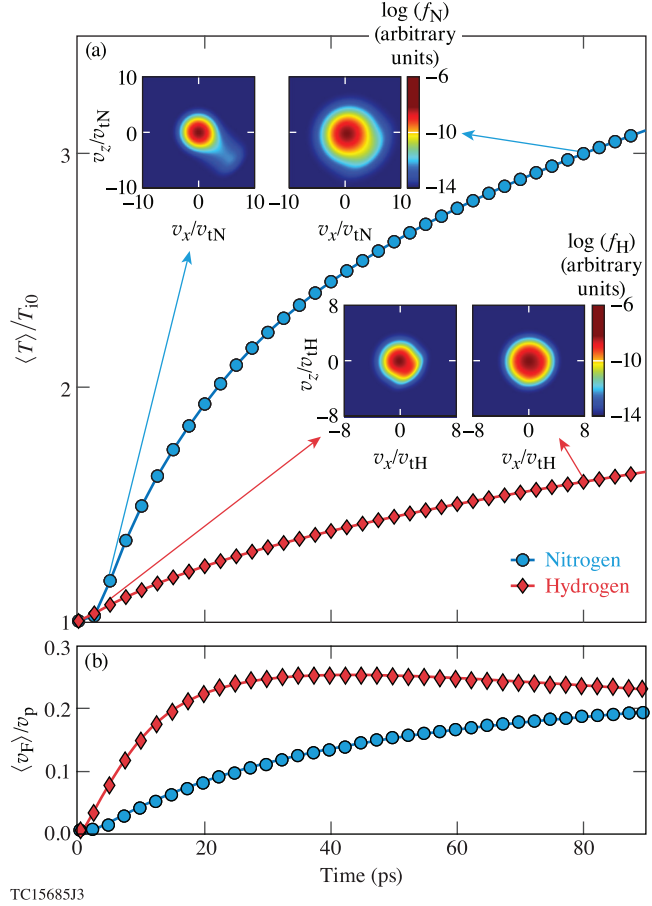}% Here is how to import EPS art
\caption{\label{fig:wide} Evolution of the (a) ion temperatures and (b) flow velocities averaged over the CBET interaction region for the case of small v$_p$. Here collisions rapidly thermalize and isotropize the tails of the distribution function (insets) that result from trapping in the IAW.}
\label{fig:f4}
\end{figure}

The rapid decay of the electrostatic energy (Fig. \ref{fig:f2}) is followed by a slow drop in the gain and marks the beginning of the late time saturation stage. During this stage, the small and large v$_p$ interactions exhibit strikingly different behavior. Foremost, the gain drops substantially for small v$_p$ and only modestly for large v$_p$. While both trends have their origin in ion trapping-induced detuning, the cause of this detuning depends on the role of each ion species in collisional energy transfer and thermalization.
Figure \ref{fig:f4}(a) displays the time evolution of the H and N temperatures averaged over the region of beam overlap ($\langle T\rangle$) for the case of small v$_p$. As observed in the experiments \cite{hansen2021cross}, the N temperature increases more rapidly and to a larger value than the H temperature. On the surface, this appears to contradict estimates based on the relative Landau damping rates that suggest the H ions should heat much more rapidly than the N ions (cf., $\partial_\mathrm{v} f_H|_{\mathrm{v}=\mathrm{v}_p}$ and $\partial_\mathrm{v} f_N|_{\mathrm{v}=\mathrm{v}_p}$) \cite{ONeilTrapping}. However, while the IAW can trap and accelerate a large number of H ions [Fig. \ref{fig:f1}(b)], these ions quickly collide with and transfer their energy to the N ions. Specifically, the H-N collision rate for the energy loss ($\epsilon$) of H ions travelling at v$_p$ is $\sim 3.5\times$ larger than the N-H rate for N ions travelling at v$_p$ ($\nu^{\mathrm{HN}}_{\epsilon} = 3.5\times 10^{10} \mathrm{s}^{-1}$ and $\nu^{\mathrm{NH}}_{\epsilon} = 1.0\times 10^{10} \mathrm{s}^{-1} $)\cite{NRLFormulary}. 
As the H transfers the energy it has gained from the IAW to the N, the N undergoes intraspecies pitch angle ($\perp$) and slowing down ($s$) collisions, relaxing its distribution to a more isotropic, thermalized shape [Fig. \ref{fig:f4}(a) insets]. Meanwhile, the H-N pitch angle and slowing down collisions isotropize the H distribution. For ions travelling at v$_p$, $\nu^{\mathrm{HH}}_{s} = 3.3\times 10^{9} \mathrm{s}^{-1}$, $\nu^{\mathrm{HH}}_{\perp} = 6.1\times 10^{9} \mathrm{s}^{-1}$,  $\nu^{\mathrm{HN}}_{\perp} = 4.7\times 10^{11}  \mathrm{s}^{-1}$, and $\nu^{\mathrm{NN}}_{s} \approx \nu^{\mathrm{NN}}_{\perp} = 1.8\times 10^{10} \mathrm{s}^{-1}$. \cite{NRLFormulary}

The isotropization of the H and N distributions occurs with respect to their mean flows in the direction of $\bold{k}$ [Fig. \ref{fig:f4}(b)]. The H flow develops during the early time saturation stage (< 20 ps) due to trapping and plateaus by the late time saturation stage (> 20 ps). The N flow, like the N temperature, evolves largely through its collisional interaction with H. While the fractional energy contained in the flow ($\sim 10\%$) is small compared to the thermal energy ($\sim 90\%$) it can, with the increase in temperature, contribute to the detuning of CBET.

\begin{figure}
\centering
\includegraphics[scale=1]{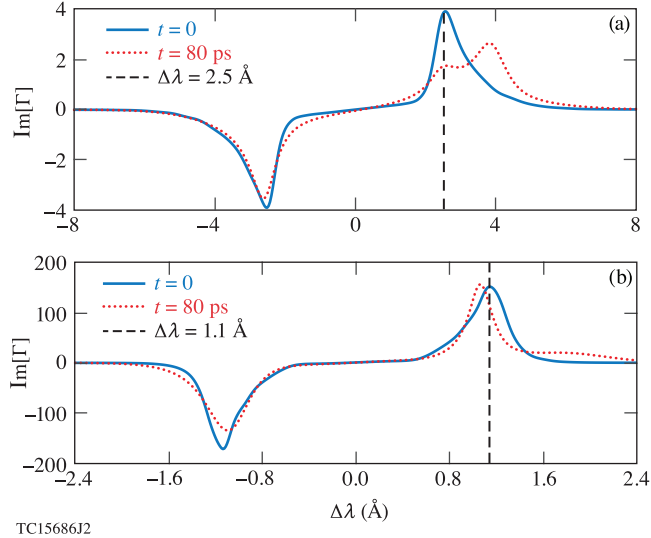}% Here is how to import EPS art
\caption{\label{fig:wide} The $\mathrm{Im}[\Gamma]$, which determines the gain in the steady state approximation [Eq. (2)], calculated using the electron and ion velocity distribution functions from VPIC at $t =$ 0 and 80 ps. Initially, the IAW is driven on resonance for both the (a) small (b) and large v$_p$ cases (intersection of the black dashed lines and the peak of the solid curve). Later, (a) the increase in ion temperatures and flow blueshift the peak in the small v$_p$ case, while (b) the persistent tails of the distribution function due to trapping redshift the peak in the large v$_p$ case. Both detunings cause the gain to drop. }
\label{fig:f5}
\end{figure}

The thermalization of the H and N ions causes a gradual blueshift in the resonant IAW frequency [Fig. \ref{fig:f5}(a)]. After $\sim 50$ ps, the plasma conditions and gain evolve slowly enough that CBET occurs in a quasi-steady state [Fig. \ref{fig:f2}(a)]. In this quasi-steady state, the kinetic coupling coefficient [Eq. (3)], calculated using the electron and ion distribution functions and averaged over the interaction region, provides the response and resonant behavior of the plasma. With the wavelength shift of the probe beam fixed at $\Delta \lambda = 2.5$ \AA, the gradual blueshift in resonant frequency causes the gain to drop. In addition, the increased damping from the modified distribution function has broadened the resonance peak and lowered its maximum. 

As in Eq. (7), both the increase in temperature and flow contribute to the blue shifting of the resonant peak. The relative shifts that each contribute can be estimated as $\Delta \lambda_{T} \sim (3\Delta T_i/ZT_e) \Delta \lambda \sim 0.4 \Delta \lambda $ and  $\Delta \lambda_{F} \sim (\mathrm{v}_F/\mathrm{v}_p) \Delta \lambda \sim 0.2 \Delta \lambda$, where $\Delta \lambda = 2.5${\AA} is the initial location of the resonance. Together, these give $\Delta \lambda_{T} + \Delta \lambda_{F} \sim 1.5${\AA}\textemdash a shift comparable to that shown in Fig. \ref{fig:f5}(a).

\begin{figure}
\centering
\includegraphics[scale=1]{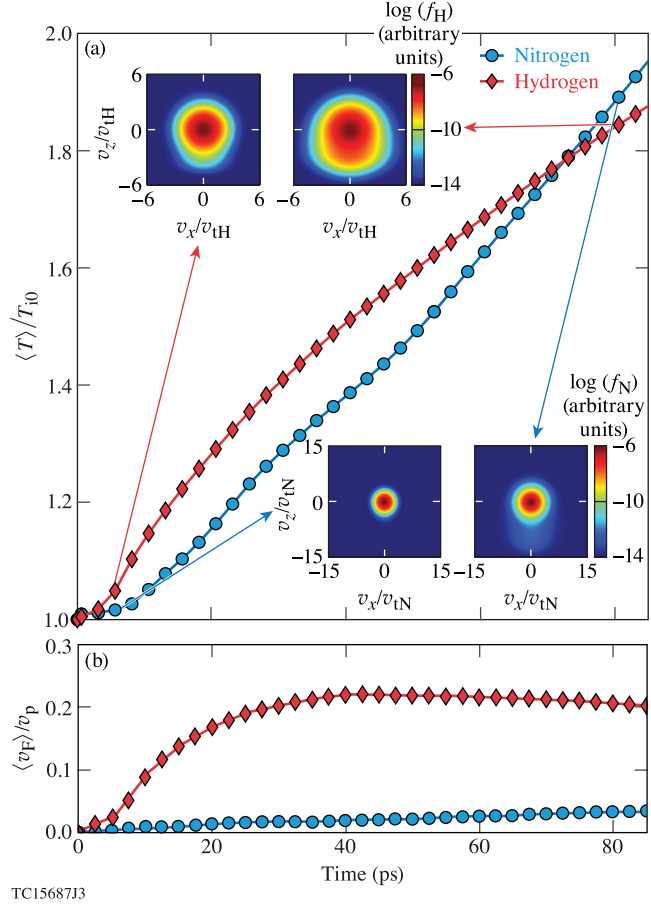}% Here is how to import EPS art
\caption{\label{fig:wide} Evolution of the (a) ion temperatures and (b) flow velocities averaged over the CBET interaction region for the case of large v$_p$. Here the lower collisionality, due to the larger velocity of the trapped ions, allows the tails of the distribution function (insets) to persist for longer times.}
\label{fig:f6}
\end{figure}

Figure \ref{fig:f6}(a) displays the time evolution of the H and N temperatures averaged over the region of beam overlap for the case of large v$_p$. Here, the H temperature increases more rapidly than the N temperature until the two eventually reach a near thermal equilibrium. As before, estimates based on the relative Landau damping rates suggest that the H should heat more rapidly than the N. In this case, however, the energy loss rates between the two species are nearly equal ($\nu^{\mathrm{HN}}_{\epsilon} \approx \nu^{\mathrm{NH}}_{\epsilon} = 1.0\times 10^{10} \mathrm{s}^{-1}$ for H and N moving at v$_p$\cite{NRLFormulary}). As a result, the H can maintain a higher temperature while it exchanges energy with the N.  

On account of the larger phase velocity, all of the collision rates are generally lower than in the small v$_p$ case ($\nu^{\mathrm{HH}}_{s} = 3.0\times 10^{9} \mathrm{s}^{-1}$, $\nu^{\mathrm{HH}}_{\perp} = 2.9\times 10^{9} \mathrm{s}^{-1}$,  $\nu^{\mathrm{HN}}_{\perp} = 1.5\times 10^{11}  \mathrm{s}^{-1}$, and $\nu^{\mathrm{NN}}_{s} \approx \nu^{\mathrm{NN}}_{\perp} = 5.2\times 10^{9} \mathrm{s}^{-1}$)\cite{NRLFormulary}. Aside from keeping the N flow velocity low [Fig. \ref{fig:f6}(b)], this allows the trapping-induced modifications to the distribution function to persist for a much longer time: even after 80 ps, both the H and N distribution functions exhibit flattening and prominent tails [Fig. \ref{fig:f6}(a) insets]. Consistent with the trapped ion frequency shift [Eq. (6)], these modifications cause a redshift in the resonant frequency [Fig. \ref{fig:f5}(b)]. 

While the pump, probe, and plasma eventually evolve to a quasi-steady state in the overlapping region ($>$50 ps in Fig. \ref{fig:f2}), the approach to this state depends on both space and time. Initially, the most intense nonlinear interaction occurs near the entrance of the pump beam into the probe beam. This region exhibits trapping, transverse breakup, and an elevated temperature before other spatial regions. The further the pump beam propagates into the probe beam, the more its intensity drops due to CBET. As a result, the processes contributing to saturation are increasingly delayed and weaker in strength further into the probe beam. Due to this spatial dependence, there is a delay in the onset of the quasi-steady state across the probe beam. 

\section{CONCLUSIONS AND OUTLOOK}
Over long time scales (>20 ps), CBET can saturate when modifications to the ion distribution functions due to trapping in the IAW create a resonance detuning. The underlying cause of this detuning depends on the phase velocity of the IAW. For small phase velocities, ion-ion collisions can rapidly thermalize the trapping-induced modification to the distribution function, such as flattening and tails. This results in an increase in the ion temperature and a residual flow that blueshifts the resonant IAW frequency. For large phase velocities, the ion-ion collision rates are lower, which allows the trapping induced modifications to persist. This results in a redshift to the resonant IAW frequency. 

The dependence on phase velocity suggests that both cases, and all cases in between, can occur in an ICF implosion, where a variety of beam intersection angles and varying plasma conditions result in a diverse spectrum of IAWs. In contrast to the situation simulated here, however, CBET in an ICF implosion involves nearly frequency degenerate beams in a flowing plasma. This flow may alter the relative importance of trapping and thermalization by transporting ions and heat out of the interaction region or by changing the saturation mechanism altogether. This will be described in a future work. 

\section*{ACKNOWLEDGEMENT}
This material is based upon work supported by the Department of Energy National Nuclear Security Administration under Award Number DE-NA0003856, the University of Rochester, and the New York State Energy Research and Development Authority.
LANL work was performed under the auspices of the U.S. Department of Energy by the Triad National Security, LLC Los Alamos National Laboratory, and was supported by the LANL Directed Research and Development (LDRD) program and the LANL Office of Experimental Science Inertial Confinement Fusion program. VPIC simulations were run on the LANL Institutional Computing Clusters.

This report was prepared as an account of work sponsored by an agency of the U.S. Government. Neither the U.S. Government nor any agency thereof, nor any of their employees, makes any warranty, express or implied, or assumes any legal liability or responsibility for the accuracy, completeness, or usefulness of any information, apparatus, product, or process disclosed, or represents that its use would not infringe privately owned rights. Reference herein to any specific commercial product, process, or service by trade name, trademark, manufacturer, or otherwise does not necessarily constitute or imply its endorsement, recommendation, or favoring by the U.S. Government or any agency thereof. The views and opinions of authors expressed herein do not necessarily state or reflect those of the U.S. Government or any agency thereof.

\appendix
\section{SUMMARY OF LINEAR CBET THEORY}
Consider two electromagnetic waves (EMWs) with wavevectors $\bold{k}_0$ and $\bold{k}_1$ and frequencies $\omega_0 > \omega_1$ crossing in a uniform density plasma devoid of flow. The waves exert a ponderomotive force on the plasma electrons, $\bold{F}_p = -\nabla \Phi_p$, where 
\begin{eqnarray}
\Phi_p = \frac{e^2}{2m_e\omega_0\omega_1}|\mathbf{E}_0+\mathbf{E}_1|^{2}
\end{eqnarray}
and $\bold{E}_0$ and $\bold{E}_1$ are the electric fields of each wave. The Coulombic attraction of the electrons and ions partially counteracts the charge separation created by the ponderomotive force. This results in a total force on the electrons and ions given by $\bold{F}_e = -\nabla (\Phi_p - e\phi)$ and $\bold{F}_s = -Z_se\nabla\phi$, respectively, where $\phi$ is the electrostatic potential and $s$ denotes the ion species.

The force associated with the frequency difference components of $\Phi_p$, i.e., those at $\Omega = \omega_0 - \omega_1$ and $\bold{k} = \bold{k}_0 - \bold{k}_1$, can strongly excite an IAW. Writing $\bold{E}_{a} = \tfrac{1}{2} \hat{\bold{E}}_ae^{i(\bold{k}_a\cdot\bold{r} - \omega_a t)} + \mathrm{c.c.}$ with $a = 0, 1$ and using this expression in Eq. (A1) provides the relevant term in the ponderomotive potential: $\Phi_{\Omega} = \tfrac{1}{2}\hat{\Phi}_{\Omega} e^{i(\bold{k}\cdot\bold{r} - \Omega t)}  + \mathrm{c.c.}$, where
\begin{eqnarray}
\hat{\Phi}_{\Omega} = \frac{e^2}{2m_e\omega_0\omega_1}\hat{\mathbf{E}}_0\cdot\hat{\mathbf{E}}_1^{*}.
\end{eqnarray}
Similarly, $\phi$ will have a component at $\Omega$ determined by the driven density perturbations, $\delta n_e$ and $\delta n_s$, and Poisson's equation:  $\phi_{\Omega} = e(\delta n_e - \sum_s{Z_s \delta n_s})/\epsilon_0 k^2$, where $\delta n_e=\tfrac{1}{2}\delta \hat{n}_ee^{i(\bold{k}\cdot\bold{r} - \Omega t)} + \mathrm{c.c.}$.

Expressions for the density perturbations can be found by using the Vlasov-Poisson system of equations. Writing the electron and ion distribution functions as
\begin{eqnarray}
f_l(\mathbf{r},\mathbf{v},t)=f_{l0}(\mathbf{v})+\tfrac{1}{2}[\delta f_{l}(\mathbf{v})e^{i(\bold{k}\cdot\bold{r} - \Omega t)} + \mathrm{c.c.}],
\end{eqnarray}
where $l = e, s$ and the $f_{l0}$ are the distribution functions in the absence of the EMWs, one can show that
\begin{eqnarray}
\frac{\delta \hat{n}_e}{n_{e0}}=-\frac{k^2}{m_e\omega_{pe}^2}\frac{\chi_e(\Omega,\mathbf{k})[1+\chi_i(\Omega,\mathbf{k})]}{1+\chi_i(\Omega,\mathbf{k})+\chi_e(\Omega,\mathbf{k})}\hat{\Phi}_{\Omega},
\end{eqnarray}
where $n_{e0}$ is the ambient electron density, $\omega_{pe} = \sqrt{e^2n_{e0}/\epsilon_0m_e}$ the associated plasma frequency, and $\chi_e$ and $\chi_i$ the electron and ion susceptibilities, respectively:
\begin{eqnarray}
\chi_e(\Omega,\mathbf{k}) = \frac{e^2}{\epsilon_0m_ek^{2}}\int \frac{\mathbf{k}\cdot \nabla_{\mathbf{v}} f_{e0}}{\Omega-\mathbf{k}\cdot\mathbf{v}}d\bold{v}
\end{eqnarray}
\begin{eqnarray}
\chi_i(\Omega,\mathbf{k}) =\sum_{s} \frac{Z_s^2e^2}{\epsilon_0m_sk^{2}}\int\frac{\mathbf{k}\cdot \nabla_{\mathbf{v}} f_{s0}}{\Omega-\mathbf{k}\cdot\mathbf{v}}d\bold{v}.
\end{eqnarray}

The electron density perturbation of the IAW ($\delta n_{e}$) and electron motion in the EMWs generate currents that couple the EMWs: $\hat{\bold{J}}_0 = -(ie/\omega_0)(n_{e0}\hat{\bold{E}}_0 +\delta \hat{n}_{e}^* \hat{\bold{E}}_1)$ and $\hat{\bold{J}}_1 = -(ie/\omega_1)(n_{e0}\hat{\bold{E}}_1 +\delta \hat{n}_{e} \hat{\bold{E}}_0)$. Using these currents in the electromagnetic wave equation for each wave and applying the paraxial and steady state approximations, one finds 
\begin{eqnarray}
(\bold{v}_a\cdot\nabla)I_a = g_{ab}I_bI_a
\end{eqnarray}
where a $\neq$ b can take the values 0 or 1, $I_{a} = \tfrac{1}{2}\epsilon_0 \mathrm{v}_a |\hat{\bold{E}}_a|^2$ is the intensity, $\bold{v}_a = (1-\omega_{pe}^2/\omega_a^2)^{1/2}c\bold{k}_a/k_a$ the group velocity, 
\begin{eqnarray}
g_{ab} = \frac{e^2k^2\mathrm{sgn}(\omega_b-\omega_a)}{2\epsilon_{0}m_{e}^2c^2\mathrm{v}_b\omega_{b}^2k_a}\mathrm{Im}
\left[\Gamma(\Omega,\bold{k})\right],
\end{eqnarray}
the gain coefficient, and
\begin{eqnarray}
\Gamma(\Omega,\mathbf{k})=\frac{\chi_e(\Omega,\mathbf{k})[1+\chi_i(\Omega,\mathbf{k})]}{1+\chi_i(\Omega,\mathbf{k})+\chi_e(\Omega,\mathbf{k})},
\end{eqnarray}
the kinetic coupling factor. For nearly all situations of interest, the frequencies $\omega_0$ and $\omega_1$ are close enough that $g_{10} \approx -g_{01} \equiv g$. 

In a 1D co-propagating geometry, the coupled equations in (A7) have the analytic solution \cite{tang1966saturation}
\begin{eqnarray}
I^{\mathrm{out}}_1 = \frac{(1+\beta)\mathrm{exp}[G_{0}(1+\beta)]}{1+\beta \mathrm{exp}[G_{0}(1+\beta)]}I^{\mathrm{in}}_1,
\end{eqnarray}
where $\beta = I^{\mathrm{in}}_1/I^{\mathrm{in}}_0$, $G_0 = gI^{\mathrm{in}}_0L$, the superscripts "in" and "out" denote the input and output values of intensity, and $L$ is the interaction length. From Eq. (A10), one can find the gain: $G = \mathrm{ln}(I^{\mathrm{out}}_1/I^{\mathrm{in}}_1)$. Note that $G$ includes the effect of pump depletion. In the limit of small $I^{\mathrm{in}}_1$, $G \to G_0$, reproducing the familiar exponential gain result, i.e, $I_1 = I^{\mathrm{in}}_1e^{G_0}$. 

As a rough estimate for the 2D or 3D gain, one can use Eq. (A10) with an effective interaction length that accounts for the crossing angle ($\theta$) and overlap of the beams: $L_{2\mathrm{D}} = d/\sin\theta$ and $L_{3\mathrm{D}} = 8d/3\pi\sin\theta$, where $d$ is the width of each beam. In both 2D and 3D, the derivation of these lengths requires flat top beams with central rays that intersect, while in 3D the beams must also have circular cross sections. For all linear gain calculations presented here, Eqs. (A10) and (A8) were used with $L = L_{2\mathrm{D}}$. In the absence of nonlinear or non-steady state effects, this approximation works surprisingly well even for speckled beams when the average intensities are used for $I^{\mathrm{in}}_0$ and $I^{\mathrm{in}}_1$. This is because the intensity transfer typically occurs over many speckles \cite{FollettCBETRay}. 

\section{VPIC SIMULATION DETAILS}
The 2D collisional VPIC simulations were performed using domains tailored to the interaction geometry. In the small v$_p$ case, the large crossing angle (99$^{\mathrm{o}})$  allowed for a smaller domain: 138 $\mu$m $\times$ 108 $\mu$m with cells of size $\Delta$x $\times$ $\Delta$z = 1.0$\lambda_D$ $\times$ 1.0$\lambda_D$. In the large v$_p$ case, on the other hand, the small crossing angle (21.4$^{\mathrm{o}})$  necessitated a larger domain: 450 $\mu$m $\times$ 154 $\mu$m with cells of size $\Delta$x $\times$ $\Delta$z = 1.2$\lambda_D$ $\times$ 1.2$\lambda_D$.  Here slightly larger cell sizes were used to recover some of the computational cost associated with the larger domain. Both cases used 512 particles per cell.

The pump and probe fields, $E_{y}$, were composed of a random distribution of f/6.7 speckles with a characteristic width and length of 2.8 and 99 $\mu$m, respectively \cite{YinSpeckles}. Absorbing boundary conditions were used for the fields.  

"Refluxing" boundary conditions were used for both the electrons and ions. Every particle that left the simulation region was replaced by a particle injected with a velocity randomly sampled from a fixed-temperature distribution. For electrons this was a super-Gaussian of order 3; for ions it was a traditional Maxwellian. 

VPIC employs a binary collision model that recovers the Landau form of inter-particle collisions for weakly coupled plasmas \cite{takizuka1977binary,yin2016plasma}. In the simulations presented here, ion-ion collisions were included, but electron-ion and electron-electron collisions were not.

\bibliographystyle{unsrt}  
\bibliography{references}  %%% Remove comment to use the external .bib file (using bibtex).


\end{document}